\documentclass[a4paper,11pt]{article}
\usepackage{jheppub} 
\usepackage[T1]{fontenc} 

\def\be{\begin{equation}}
\def\ee{\end{equation}}
\def\bea{\begin{array}}
\def\eea{\end{array}}
\def\beqa{\begin{eqnarray}}
\def\eeqa{\end{eqnarray}}
\def\beqas{\begin{eqnarray*}}
\def\eeqas{\end{eqnarray*}}
\def\ol{\overline}
\def\bp{\begin{picture}}
\def\ep{\end{picture}}
\def\bc{\begin{center}}
\def\ec{\end{center}}
\def\bfig{\begin{figure}}

\def\efig{\end{figure}}

\def\bit{\begin{itemize}}
\def\eit{\end{itemize}}
\def\nn{\nonumber}
\def\f{\frac}

\def\[{\left[}
\def\]{\right]}
\def\({\left(}
\def\){\right)}
\def\l.{\left.}
\def\.{\right.}
\def\l|{\left|}
\def\r|{\right|}
\def\tl{\tilde}
\def\ra{\rightarrow}
\def\la{\leftarrow}

\def\tm{\times}

\def\s{\rm s}

\def\da{\dagger}

\def\la{\lambda}

\def\al{\alpha}

\def\ka{\kappa}

\def\ep{\epsilon}

\def\pa{\partial}
\def\pr{\prime}

\title{\boldmath Top and Bottom Seesaw from Supersymmetric Strong Dynamics}

\author[a]{Csaba Balazs,}
\author[b,c,d]{Tianjun Li,}
\author[e,1]{Fei Wang,\note{Corresponding author.}}
\author[b,c]{Jin Min Yang}

\affiliation[a]{School of Physics, Monash University, Melbourne Victoria 3800,
Australia}
\affiliation[b]{Kavli Institute for Theoretical Physics China (KITPC),
Institute of Theoretical Physics, Chinese Academy of Sciences,
Beijing 100190, P. R. China}
\affiliation[c]{State Key Laboratory of Theoretical Physics,
Institute of Theoretical Physics, Chinese Academy of Sciences,
Beijing 100190, P. R. China}
\affiliation[d]{George P. and Cynthia W. Mitchell Institute for
Fundamental Physics and Astronomy, Texas A$\&$M University,
College Station, TX 77843, USA}
\affiliation[e]{School of Physics, Zhengzhou University, 450000, Zhengzhou
P. R. China}

\emailAdd{Csaba.Balazs@sci.Monash.edu.au}
\emailAdd{tli@itp.ac.cn}
\emailAdd{feiwang@zzu.edu.cn}
\emailAdd{jmyang@itp.ac.cn}

\abstract{ We propose a top and bottom seesaw model with partial composite top and bottom quarks.
Such composite quarks and topcolor gauge bosons are bound states from supersymmetric
strong dynamics by Seiberg duality. Supersymmetry breaking also induces the breaking
of topcolor into the QCD gauge coupling. The low energy description of our model reduces
to a complete non-minimal extension of the top seesaw model with bottom seesaw. The non-minimal
nature is crucial for Higgs mixings and the appearance of light Higgs fields. The Higgs
fields are bound states of partial composite particles with the lightest one compatible
with a 125 GeV Higgs field which was discovered at the LHC.
}

\begin{document}
\maketitle
\flushbottom

\newpage
\section{Introduction}
The Standard Model (SM) of electroweak interactions, based on the spontaneously
$SU(2)_L \times U(1)_Y$ gauge symmetry breaking, has
been extremely successful in describing phenomena below the electroweak scale.
The most important problem in the SM is the source of the electroweak gauge
symmetry breaking and the related problem of hierarchical flavor structure.
It is well known that the top quark is very heavy comparing to the other SM fermions
and its value is obtained in an ad hoc manner by adjusting the phenomenologically
introduced Yukawa couplings. Besides, the top quark couples more strongly to
electroweak symmetry breaking sector than the light quarks and it is possible
that some of the electroweak symmetry breaking is due to top sector. The idea of
top condensation~\cite{topcondensation} is an attractive approach to explain these problems.

However, the minimal top condensation framework predicts a too high top quark mass
$m_t$ as well as
a high Higgs mass, and then the extreme fine-tuning is needed to trigger the condensation.
Also, the Nambu-Jona-Lasinio (NJL) model must be considered as an approximation of some new
strong dynamics--the topcolor gauge interactions. One can combine topcolor with technicolor
to get a TC2 model~\cite{tc2} in which the electroweak symmetry breaking gets contributions
from both the top condensation and the technicolor sectors. The other very interesting
scenario is the top seesaw model~\cite{topseesaw} which naturally predicts the acceptable
top quark mass without the
need of new electroweak symmetry breaking sector. The UV completion of topcolor needs more
matter contents and certain interactions which are put in by hand. We would like to
present a model which will give rise to these terms automatically.

It is well known that the SM requires the existence of Higgs fields to trigger electroweak gauge
symmetry breaking. However, the quantum corrections to Higgs boson masses have quadratic divergences.
Thus, the entire SM mass spectrum, which depends on the Vacuum Expectation
Value (VEV) of Higgs field, is directly and indirectly sensitive to the cut-off scale
of the theory like the Planck scale. This is the gauge hiearchy problem.
One natural solution is supersymmetry (SUSY)
by adding supersymmetric partners of the SM particles to cancel the quadratic divergences.
However, the ATLAS and CMS Collaborations at the LHC have not found any signal of supersymmetric
particles (sparticles) yet. Moreover, SUSY can provide a viable dark matter candidate,
achieve the gauge coupling unification as well as be an essential ingredient to certain
quantum gravity candidate. Thus, it is possible that our Universe could adopt
supersymmetry at relatively high scale.

It had been conjectured long time ago that all the building blocks of the SM are
composite particles instead of being fundamental particles.
The existence of chiral symmetry is essential to guarantee the disappearance of the
known fermion masses. However, 't~Hooft anomaly matching conditions~\cite{thooft} are very
restrictive and hardly can one obtain the realistic composite models. A very interesting
progress was achieved by Seiberg who discovered the duality~\cite{seibergduality} between
different SUSY gauge theories. Seiberg duality is highly non-trivial and satisfies the 't~Hooft
anomaly matching conditions and decoupling conditions as well as the other consistent checks.
Besides, new emergent gauge groups and composite fermions appear in certain case of
the dual description. We conjecture that the SM particles are composite
and such compositeness are the consequences of SUSY strong dynamics and SUSY breaking.
The observed small mass terms of the SM fermions are the consequences of the strong
dynamics arise from the emergent gauge interactions. Especially, the Higgs boson
mass around 125~GeV, which was discovered at the LHC recently~\cite{:2012gk, :2012gu},
can be realized as well.

This paper is organized as follows. In Section~\ref{sec-1}, we
discuss the emergent topcolor gauge group and matter contents from SUSY strong dynamics.
SUSY is broken by rank conditions in our scenario, which results in the ISS-type metastable
vacua \cite{ISS}.
In Section~\ref{sec-2}, we discuss the complete top and bottom seesaw sector.
The composite matter content from Seiberg duality results in partial composite
physical top and bottom quarks. Composite multiple Higgs doublets appear in our model
at low energy and are fully responsible for electroweak gauge symmetry breaking.
Section~\ref{sec-3} contains our Conclusions.

\section{Composite Particles from SUSY Strong Dynamics}
\label{sec-1}
Top quark, which couples more strongly to the electroweak symmetry breaking sector
than other light quarks, could be responsible for electroweak symmetry breaking.
The idea of top condensation is fairly attractive and gives an explanation on how top quark
can participate in the electroweak symmetry breaking mechanism and obtain a
dynamically-generated mass term. The UV completion of the top condensation idea suggests
the existence of new topcolor gauge interactions. The complete topcolor sector requires
new Higgs fields in $({\bf 3,\bar{3}})$ representation to break the topcolor
gauge symmetry down to $SU(3)_C$.
Besides, top seesaw sector requires new vector-like particles. We want to obtain all
the required ingredients from SUSY strong dynamics. The most simple setting is
the vector-like supersymmetric QCD.

Let us consider $SU(N_C)$ SUSY QCD which has the massive vector-like quarks
$Q_i$ and $\tl{Q}^i$ with $i=1,~2,~...,~N_F$, and several $SU(N_C)$ singlet massive
messenger fields $\bar{f}_k$ and $f_k (k=1,...,n_I)$ for gauge mediation.
The global flavor symmetry is
$SU(N_F)_1\times SU(N_F)_2\times U(1)_V\times U(1)_R$. We adopt the following superpotential
\beqa
W= m^{i}_jQ_i\tl{Q}^j+\ka^{ij}\f{Q_i\tl{Q}^j\bar{f}^kf_k}{M_*}+M_0 \bar{f}^kf_k.
\eeqa
where the following mass terms
\beqa
m^i_j=m_0 \delta_j^i~,~
\eeqa
break the flavor symmetry down to $SU(N_F)_V\times U(1)_V$ and $M_*$ some new mass scale below which non-renormalizable operators of the form in the formula is generated.
This superpotential is of the simplified gauge mediation type proposed in Ref.~\cite{murayama}.

According to the Seiberg duality~\cite{seibergduality}, this theory is dual to
an $SU(N_F-N_C)$ gauge theory.
We can identify the dual magnetic gauge group as the new topcolor-like $SU(3)_1$.
Besides, we require that the dual magnetic gauge group be IR-free which sets $N_C+1<N_F<3/2 N_C$.
Thus, the only possible choice is $N_C \ge 6$. We chose $N_C=7$ and $N_F=10$ in our scenario.

We also embed the gauge symmetry $SU(3)_2\times SU(2)_L\times U(1)_Y$ into the $SU(6)$ subgroup of
the global symmetry $SU(10)_V$ by assigning
\beqa
{Q}^T&=&( 3, 1)_{0},~~ {Q}^D=(1, 2)_{-1/3},~~{Q}^S=(1,1)_{2/3},\nn\\~~ \tl{Q}^T&=&(\bar{3},1)_{0},~~ \tl{Q}^D=(1,2)_{1/3},~~ \tl{Q}^S=(1,1)_{-2/3}~.
\eeqa
We also embedding an additional $U(1)_1$ into $U(1)_V$. The purpose of such additional $U(1)_1$
will be clear later. The fields $Q^T$ and $\tl{Q}^T$, etc, are gauge singlets with respect to $U(1)_1$. However, the messenger fields $f_k$ and $\bar{f}^k$ carry non-zero $U(1)_1$ charge.

The electric theory is dual to a magnetic $SU(3)_1$ gauge theory with superpotential
\beqa
W=h Tr ( \tl{q}\tl{M} q) + h \Lambda m_0 Tr (\tl{M}) + \f{\Lambda}{M_*}Tr(\ka\tl{M})\bar{f}^kf_k + M_0 \bar{f}^kf_k~,~
\eeqa
and the scale is defined as
\beqa
(-1)^{N_f-N_c}\Lambda^{b_e+b_m}=\Lambda_e^{3N_c-N_f}{\Lambda}_m^{2N_f-3N_c}~,
\eeqa
where $b_e$ and $b_m$ are respectively the SUSY QCD beta functions of the electric and magnetic theories
with the respectively dynamical transmutation scales $\Lambda_e$ and $\Lambda_m$.

In general, the SUSY breaking requires the presence of R-symmetry~\cite{rsusy}. However,
an exact R-symmetry forbids gaugino masses which is not acceptable. One possible solution
is to explicitly break the R-symmetry by introducing small R-symmetry violation terms
which lead to meta-stable vacua. In our scenario, we can see that the first three terms
have a $U(1)_R$ symmetry with $R(\tl{M})=2$ and $R(\tl{q})=R(q)=R(\bar{f})=R(f)=0$.
Such an exact $U(1)_R$ symmetry is obviously broken to an approximate one by the last term.

The magnetic theory requires the existence of meson-like composites to satisfy the anomaly
matching conditions. Components of the meson fields $\tl{M}$ from $Q^T,Q^D,Q^S$ and
$\tl{Q}^T,\tl{Q}^D,\tl{Q}_S$ composition can be decomposed in terms of
$SU(3)_2\times SU(2)_L\times U(1)_Y$ as
\beqa
\tl{Q}^TQ^T&\sim& (8,1)_{0}\oplus (1,1)_{0},\nn\\
\tl{Q}^TQ^D\oplus \tl{Q}^DQ^T&\sim& (3,2)_{1/3}\oplus (\bar{3},2)_{-1/3},\nn\\
\tl{Q}^TQ^S\oplus \tl{Q}^SQ^T&\sim& (3,1)_{-2/3} \oplus(\bar{3},1)_{2/3} ,\nn\\
\tl{Q}^DQ^S\oplus \tl{Q}^SQ^D&\sim& (1,2)_{-1} \oplus(1,2)_{1},\nn\\
\tl{Q}^DQ^D\oplus\tl{Q}^SQ^S&\sim&(1,3)_{0}\oplus(1,1)_{0}\oplus(1,1)_{0}.
\eeqa
Similarly, the $({3},\bar{6})/(\bar{3},{6})$ components of the dual quarks $({3},\overline{10})/(\bar{3},{10})$ are transformed in terms of $SU(3)_1\times SU(3)_2\times SU(2)_L\times U(1)_Y$ as
\beqas
{q}({3},\bar{6})&\sim& ({3},\bar{3},1)_{0}\oplus({3},1,2)_{1/3}\oplus({3},1,1)_{-2/3}~,\nn\\
\tl{q}(\bar{3},{6})&\sim& (\bar{3},3,1)_{0}\oplus (\bar{3},1,2)_{-1/3}\oplus(\bar{3},1,1)_{2/3}~.
\eeqas
Thus, in our theory we can identify
\beqa
&&T_L\equiv\(\bea{c} t_L \\ b_L \eea \)\sim (3,1,{2})_{1/3},~~
X_L^c\equiv\(\bea{c}\chi_L^c\\ \omega_L^c \eea\)\sim (\bar{3},1,{2})_{-1/3},~~
X_L\equiv\(\bea{c}\chi_L\\ \omega_L \eea\)\sim (1, {3},{2})_{1/3}~,\nn\\
&&
P_L^c\equiv~\(\bea{c}\rho_L^c \\ \sigma_L^c\eea\)\sim (1, \bar{3},{2})_{-1/3},~ b_L^c\sim(1,\bar{3},1)_{2/3}~,~~
\tl{\omega}_L\sim (1,3,1)_{-2/3}~,~\tl{\sigma}_L\sim (3,1,1)_{-2/3},~\nn\\
&&\tl{\omega}_L^c \sim (\bar{3},1,1)_{2/3},~\tl{\sigma}_L^c~\sim (1,\bar{3},1)_{2/3},~H_1\sim(1,1,2)_{-1},~
H_2\sim (1,1,2)_{1},~\nn\\
&&\Phi_1 \sim (3,\bar{3},1)_{0},~\Phi_2\sim (\bar{3},3, 1)_{0},~S^a \sim (1,1,1)_{0}^a~(a=1,2)~.
\eeqa

 From the dynamical superpotential by Seiberg duality, we can identify the following interactions
\beqa
W&\supset& h \(\bea{c}\chi_L\\ \omega_L \eea\)^T \Phi_1 \(\bea{c}\chi_L^c\\ \omega_L^c \eea\)+h \(\bea{c}t_L\\ b_L \eea\)^T \Phi_2 \(\bea{c}\rho_L^c\\ \sigma_L^c \eea\)
+h \(\bea{c}t_L\\ b_L \eea\)^T \(\bea{c}\chi_L^c\\ \omega_L^c \eea\) S_a\nn\\
&+&h\tl{\omega}_L^c \Phi_1\tl{\omega}_L +h\tl{\sigma}_L^c \Phi_2\tl{\sigma}_L+ h \(\bea{c}t_L\\b_L\eea\)H_1\tl{\omega}_L^c+h\(\bea{c}\chi_L^c\\\omega_L^c\eea\)H_2\tl{\sigma}_L
+h\tl{\omega}_L\tl{\sigma}_L^c S_a~.~\,
\eeqa

We also introduce the right-handed top quark chiral supermultiplets in terms of gauge group
$SU(3)_2\times SU(2)_L\times U(1)_Y\times U(1)_1$ quantum number
\beqa
t_L^c\sim (1,\bar{3},1)_{(-4/3, 1)}~,~~~P_L\equiv(\rho_L,\sigma_L)\sim (1,3,2)_{(1/3,1)},
\eeqa
and possible Higgs sector to completely break $U(1)_1$ at low energy. The necessity of chiral fermions is
obvious. SUSY QCD is vector-like and the resulting dual gauge theory is still vector-like. In order to
get the chiral fermions, we must introduce by hand at least one chiral component. This fact also
appears in the (latticed) extra dimensional interpretation of top seesaw~\cite{latticetop}.
Localized heavy kink mass terms are necessary to get the localized chiral fermions.

Supersymmetry is broken by rank conditions~\cite{ISS}. Neglecting temporarily the contributions
of the messenger fields, we can see from the rank conditions
\beqa
-F^*_{\tl{M}_i^j}=\la \tl{q}^iq_j + \Lambda\delta_{i}^j m_0~,
\eeqa
that supersymmetry is indeed broken. This is a typical in ISS-like models.
The scalar potential is minimized along a classical pseudo-moduli space
of vacua which is given by~\cite{ISS}
\beqa
M=\(\bea{cc}0~&~0\\0~&~\phi_0\eea\)~,{q}=\(\bea{c} q_0\\0\eea\)~,\tl{q}=\(\bea{c} \tl{q}_0\\0\eea\)~,
\eeqa
with
\beqa
q_0\tl{q}_0=m_0\Lambda~,
\eeqa
and arbitrary $\phi_0$. In our scenario, the $q_0$ and $\tl{q}_0$ parts corresponding to the
VEVs of $\Phi_1$ and $\Phi_2$ fields within the dual quark decomposition.

Flat pseudo-moduli will in general be lifted by quantum effects. The one-loop stable minimum
by Coleman-Weinberg potential~\cite{ISS} is
\beqa
\phi_0= {\bf 0}_{\tl{N}\times \tl{N}}~,~q_0=M_1{\bf 1}_{N\tm N}~,\tl{q}_0=M_2{\bf 1}_{N\tm N}~, M_1=M_2=\sqrt{-m_0\Lambda}~.
\eeqa
The $U(1)_R$ violation terms involving the messengers will shift the minimum of $\tl{M}$ through one-loop Coleman-Weinberg potential by an amount
\beqa
<\phi_0>=\Delta \phi_0 \equiv s_1 \sim \f{\la^3 m_0 \Lambda^4}{M M_{*}^3}~.
\eeqa
The lifetime of the metastable vacua requires
\beqa
|\epsilon|\sim \sqrt{\f{m_0}{\Lambda_m}}\ll1~,
\eeqa
with the tunneling probability $e^S$ to exceed the lifetime of our universe $e^{40}$ seconds
\beqa
S\sim \epsilon^{-\f{4(3N_c-N_f)}{N_f-N_c}}>40.
\eeqa
There are large viable parameter spaces that can satisfy this requirement.

Possible new SUSY breaking minimum can arise through the combination of $m_{ij}$ and
$\f{\ka_{ij}\Lambda}{M_*}\bar{f}^kf_k$. For example, a possible new minimum may be possible
if $\bar{f}^kf_k=\f{mM_{*}}{\ka}$. However, the lifetime (for tunneling to the new possible minimum) of the previous metastable vacuum can be
long enough if we set
\beqa
\f{M^2 M_*}{\la}\gtrsim m\Lambda^2~.
\eeqa

The F-term of the meson fields induce the scalar mass terms for
$T_L,~X_L^c,~\tl{\sigma}_L,~{\rm and}~\tl{\omega}_L^c$ from the induced superpotential
$\tl{q}M q$. Other soft SUSY masses can be generated through the effective messenger fields
\beqa
M_{mess}=M_0+\f{\ka \Lambda}{M_{*}}<\tl{M}> \simeq M_0~,
\eeqa
with
\beqa
F_{mess}=\ka_{ij}\Lambda\f{F_{\tl{M}_i^j}}{M_*}=\f{\ka_{ii} m_0\Lambda^2}{M_*}~.~\,
\eeqa
Thus, we obtain the gaugino masses~\cite{GMSB}
\beqa
M_{a}\simeq \f{\al_a}{4\pi} \sum\limits_{I}n_a(I)\f{ F_M}{M},
\eeqa
and sfermion masses
\beqa
m_{\phi_i}^2\simeq 2\[\sum\limits_{a}\(\f{\al_a}{4\pi}\)^2 C_a(i)n_a(I)\]\(\f{F_M}{M}\)^2~.
\eeqa
Below the scale $\sqrt{F}$ which is the typical scalar masses for dual squarks, the SUSY QCD reduce to non-supersymmetric dynamics.
The gaugino and remaining sfermions can acquire masses from gauge loops. The matter contents participate in (part of) the following types of interactions $SU(3)_1\times SU(3)_2\times SU(2)_L\times U(1)_Y\times U(1)_1$. Besides, the soft masses of remaining superpartner are controlled by the messenger mass parameter $M$ as well as $F_M$. We will see shortly that the additional $U(1)_1$ coupling as well as one SU(3) is nearly strong-coupled, thus dominate the gauge mediation contributions to the soft sfermion masses. Requiring the scale $M$ and $\sqrt{F_M}$ is comparable to each other and taking into account the messenger species multiplication factor $n_a(I)$, we can easily tune the soft squark and gaugino masses to lie near $\sqrt{F}$.
 Thus, after integrating out the relevant supersymmetry partners, we get at the low energy an $SU(3)_1\times SU(3)_2\times SU(2)_L\times U(1)_Y\times U(1)_1$ gauge theory with proper matter contents and interactions. The gauge group of the $SU(3)_1$ is emergent and almost all the matter contents are composite particles.

\section{Top and Bottom Seesaw}
\label{sec-2}
It is well known from the topcolor dynamics that the predicted top quark mass is too high if
topcolor is responsible for full electroweak symmetry breaking. In order to get realistic
top quark mass, top seesaw model was proposed by introducing additional vector-like particles
besides the topcolor matter content. In our model, partial composite top and bottom quarks
will naturally lead to top and bottom seesaw mechanism.

After $\Phi_1$ and $\Phi_2$ respectively acquire VEVs
$M_1$ and $M_2$, the $SU(3)_1\times SU(3)_2$ gauge symmetry is broken down
to $SU(3)_C$.
The theory has a set of massless gluons and massive octet colorons. The remaining QCD coupling is
\beqa
\f{1}{g_c^2}&=&\f{1}{h_1^2}+\f{1}{h_2^2}~,~~~~~\cot\theta=\f{h_1}{h_2}~,
\eeqa
where $h_1$ and $h_2$ are the gauge couplings for $SU(3)_1$ and $SU(3)_2$,
and the massive colorons acquire masses $M_B^2=(h_1^2+h_2^2) (M_1^2+M_2^2)$.

After we integrate out the coloron fields and the sfermions for the third generation quarks, we obtain the effective four-fermion interactions
\beqa
{\cal L}&=&{\cal L}_{\rm kinetic}-(M_2 X_L^T C X_L^c+ M_1 T_L^T C P_L^c+ s_1 T_L^T C X_L^c )+{\cal L}_{\rm int}
\eeqa
with
\beqa
{\cal L}_{\rm int}=-\f{h_2^2}{M_B^2}\({X}_L^\dagger\bar{\sigma}^\mu \f{M^A}{2}X_L\)\({t}_L^c\sigma^\mu\f{M^A}{2}(t_L^c)^\dagger\)+(X_L\ra P_L,t_L^c\ra P_L^c )+\cdots.
\eeqa
After performing the Fierz rearrangement and at the leading order in $1/N_c$, we have
\beqa
{\cal L}_{int}&=&\f{h_2^2}{M_B^2}\[\f{}{}(\bar{X}_L t_R)(\bar{t}_R X_L)+(\bar{X}_L P_R)(\bar{P}_R X_L)+
(\bar{P}_L P_R)(\bar{P}_R P_L)+\cdots \]~.
\eeqa
We can transform the interaction eigenstates to the partial mass eigenstates by
\beqa
t_L^c\ra t_L^{c\pr}=t_L^c~,~~ X_L^c &\ra& X_L^{c\pr}= X_L^c\cos\beta+P_L^c\sin\beta~,~~P_L^c\ra P_L^{c\pr}=-X_L^c\sin\beta+P_L^c\cos\beta~,\nn\\
\eeqa
where
\beqa
~\tan(2\beta)=\f{2s_1 M_1}{s_1^2+M_2^2-M_1^2}~.
\eeqa
We define the unitary mixing matrix $\tl{T}_i \equiv N_{ij}^{-1}T_j$ with
\beqa
\tl{T}_1\equiv T_L^\pr,~\tl{T}_2\equiv X_L^\pr,~\tl{T}_3\equiv P_L^\pr,~T_1\equiv T_L,~T_2\equiv X_L,~T_3\equiv P_L~.\eeqa
In this basis, the NJL model takes the form
\beqa\small
{\cal L}&=&{\cal L}_{\rm kinetic}-(\overline{M}_1\overline{X}_L^\pr X_R^{\pr}+\overline{M}_2\overline{P}_L^\pr P_R^{\pr})
+\f{h_2^2}{M_B^2}\left\{\f{}{}\[\(\sum\limits_{i=1}^3 N_{2i}\ol{\tl{T}}_{i}\)t_R^\pr\]\[\bar{t}_R^\pr\(\sum\limits_{i=1}^3 N_{2i}\ol{\tl{T}}_{i}\)\]\right.\nn\\
&+&\left.\[\(\sum\limits_{i=1}^3\sum\limits_{j=2}^3 N_{ji}\ol{\tl{T}}_{i}\)\(-X_R^\pr\sin\beta+P_R^\pr\cos\beta\)\] \[\(-\bar{X}_R^\pr\sin\beta+\bar{P}_R^\pr\cos\beta\)\(\sum\limits_{i=1}^3\sum\limits_{j=2}^3 N_{ji}{\tl{T}}_{i}\)\]\right\}\nn\\
\eeqa
with $\overline{M}_1$ and $\overline{M}_2$ the eigenvalues of the matrix
\beqa
\(\bea{cc}s_1^2+M_2^2&~s_1M_1\\s_1M_1&~M_1^2\eea\)~.
\eeqa

Assume the gauge couplings for $SU(3)_2$ and $U(1)_1$ get strong quickly towards IR and
trigger the fermion condensation.
The vacuum is tilted by the $U(1)_1$ interactions so that condensation between
$\rho_L$ and $t_L^c$ is disallowed by the repulsive forces of $U(1)_1$.
 From the expansion, we can see that possible types of dynamical condensations for $\bar{X}_L t_R$ are
\beqa
<\bar{t}_L^\pr t_R^\pr>,~~<\bar{\chi}_L^\pr t_R^\pr>,~~<\bar{\rho}^\pr_L t_R^\pr>,
\eeqa
with corresponding mass gap
\beqa
-\mu_{tt}\bar{t}_L^\pr t_R^\pr-\mu_{\chi t}\bar{\chi}_L^\pr t_R^\pr-\mu_{\rho t} \bar{\rho}_L^\pr t_R^\pr~.
\eeqa
And they have the following relations
\beqa
\mu_{tt}=\mu N_{21},~~\mu_{t\chi}=\mu N_{22},~~\mu_{t\rho}=\mu N_{23}~,
\eeqa
so that they are not independent.
Just as the case for ordinary topcolor, the dynamical mass terms $\mu$ can be
calculated through the gap equations. The relevant diagrams are shown in Fig(1).
Detailed expressions for $\bar{\chi}_L^\pr t_R^\pr$ condensation can be seen in appendix A.
From the gap equation, we can get the analytical expressions for the condensation
scale $\mu$ and the effective critical coupling. This approach with mass insertion
is an approximation at large $N_c$ expansion. We will deduce more precise forms
of the condensation in symmetry broken phase.

Similarly, we can get the other condensations
\beqa
<\bar{t}_L^\pr \rho_R^\pr>,<\bar{\chi}_L^\pr \rho_R^\pr>,<\bar{\rho}^\pr \rho_R^\pr>, \cdots
\eeqa
to give $<\bar{X}_L P_R>$ and $<\bar{P}_L P_R>$.
After all condensation occurs, we get the most general possible mass matrix for top sector
\beqa
(t_L~, \chi_L~, \rho_L)\(\bea{ccc}0&s_a&M_1\\ \mu &M_2&\mu_1 \\ 0&0&\mu_2 \eea\)\(\bea{c}t_L^c\\ \chi_L^c\\ \rho_L^c \eea
\)~.\eeqa
The mass eigenvalues and eigenstates can be obtained by the following unitary transformations
\beqa
{\cal M}=U_L^\da{\cal M}_{\rm diag} U_R.
\eeqa
The analytical expressions are very complicate. Careful analysis indicates that the three mass eigenvalues are of order
\beqa
({m}_\chi)_{Phy}^2{\equiv}\la_2^2\sim M_1^2,~({m}_{\rho})_{Phy}^2{\equiv}\la_3^2\sim M_2^2,~(m_t)_{Phy}^2{\equiv}\la_1^2\approx \f{s_a^2 \mu^2\mu_2^2}{M_1^2M_2^2},
\eeqa
in case $M_2=M_1\gtrsim s_a\gg\mu$.
We will not give the explicit expression of the mass eigenstates
for the top quark sector. We just parameterize them as
\beqa
(t_L^m, \chi_L^m, \rho_L^m)^T={U}_{ij}^L (t_L, \chi_L, \rho_L)^T~,~(t_R^m, \chi_R^m,\rho_R^m)^T= {U}^R_{ij} (t_L, \chi_L, \rho_L)^T~,
\eeqa
with the lowest mass eigenstates $t^m_{L,R}$ corresponding to the physical top quark.
One Higgs doublet field in the multiple-Higgs-doublets are the condensations
\beqa
H_1\sim (\bar{\chi}_Lt_R,\bar{\omega}_Lt_R)=((h_0+\pi_t^0+v_{h_0})/\sqrt{2},\pi_t^+~)~,
\eeqa
and additional two singlets (and triplets) are from the condensations
\beqa
H_2\sim \bar{X}_L\otimes P_R=\Delta_2({\bf 3})\oplus S_2({\bf 1})
~,H_3\sim \bar{P}_L\otimes P_R=\Delta_3({\bf 3})\oplus S_3({\bf 1})~.
\eeqa
We obtain the precise gap equation of this theory in the
broken phase~\cite{hongjian} at the cut-off scale $M$
\beqa
{\cal L}_{\Lambda}&=&-(\ol{t}_L~, \ol{\chi}_L~, \ol{\rho}_L)\(\bea{ccc}0&s_a&M_1\\ \mu &M_2&\mu_1 \\ 0&0&\mu_2 \eea\)\(\bea{c}t_R\\ \chi_R\\ \rho_R \eea\)
-\f{h_2 M}{\sqrt{2}M_B}\bar{\chi} t (h_0+v_{h_0})
-\f{ih_2 M}{\sqrt{2}M_B}\bar{\chi}\gamma^5 t \pi_t^0\nn\\
&&-\f{h_2 M}{M_B}\bar{\omega}_Lt_R\pi_t^- - \f{h_2 M}{\sqrt{2}M_B} \bar{X}_LP_R (S_2+\Delta_2+v_{S_1})
-\f{h_2 M}{\sqrt{2}M_B} \bar{P}_LP_R (S_3+\Delta_3+v_{S_2})\nn\\
&&
-\f{1}{2} M^2\(h_0^2+\sum\limits_{i=2}^3(S_i^2+2|\Delta_i|^2)\)-M^2 \(v_{h_0} h_0+\sum\limits_{i}v_{S_i}S_i\)\nn\\
&\supseteq&-\la_1\bar{t}^m_L t^m_R-\la_2\bar{\chi}^m_L\chi^m_R-\la_3\bar{\rho}_L^m\rho_R^m \nn\\
&& -\f{1}{2} M^2\(h_0^2+\sum\limits_{i=2}^3(S_i^2+2|\Delta_i|^2)\)-M^2 \(v_{h_0} h_0+\sum\limits_{i}v_{S_i}S_i\)\nn\\
&&-\f{h_2 M}{\sqrt{2}M_B}\(U_{L21}^{-1}\bar{t}^m_L+U_{L22}^{-1}\bar{\chi}^m_L+U_{L23}^{-1}\bar{\rho}^m_L\)\(U_{R11}^{-1}{t}^m_R+U_{R12}^{-1} {\chi}^m_R+U_{R13}^{-1}\bar{\rho}^m_R\)h_0\nn\\
&&-\f{h_2 M}{\sqrt{2}M_B}\(U_{L21}^{-1}\bar{t}^m_L+U_{L22}^{-1}\bar{\chi}^m_L+U_{L23}^{-1}\bar{\rho}^m_L\)\(U_{R31}^{-1}{t}^m_R+U_{R32}^{-1} {\chi}^m_R+U_{R33}^{-1}\bar{\rho}^m_R\){S}_2\nn\\
&&-\f{h_2 M}{\sqrt{2}M_B}\(U_{L31}^{-1}\bar{t}^m_L+U_{L32}^{-1}\bar{\chi}^m_L+U_{L33}^{-1}\bar{\rho}^m_L\)\(U_{R31}^{-1}{t}^m_R+U_{R32}^{-1} {\chi}^m_R+U_{R33}^{-1}\bar{\rho}^m_R\){S}_3.
\eeqa
After we integrate out the heavy fields $\chi^m,\rho^m$, we obtain the low energy effective theory
\beqa
{\cal L}_{\mu}&=&-\la_1\bar{t}^m_L t^m_R+ \f{h_2 M}{\sqrt{2}M_B}U_{L21}^{-1}\bar{t}^m_LU_{R11}^{-1}{t}^m_Rh_0+ \f{h_2 M}{\sqrt{2}M_B} U_{L21}^{-1}\bar{t}^m_LU_{R31}^{-1}{t}^m_R\tl{S}_2\nn\\
&&+\f{h_2 M}{\sqrt{2}M_B}U_{L31}^{-1}\bar{t}^m_LU_{R31}^{-1}{t}^m_R\tl{S}_3
+\f{1}{2}Z_{h_0}(\pa_\mu h_0)^2+\f{1}{2}\sum\limits_{i}Z_{S_i}(\pa_\mu S_i)^2
+\sum\limits_{i}Z_{\Delta_i}|\pa_\mu\Delta_{i}|^2 \nn\\
&&-\f{1}{2}\(M_{h_0}^2h_0^2+\sum\limits_{i}M_{S_i}^2 S_i^2\)-\sum\limits_{i}M_{\Delta_i}^2 |\Delta_i^2|^2-\sum\limits_{i}M_{i0} S_i h_0\nn\\
&&-\tl{M}_{23}S_2S_3-M_{23}\Delta_2\Delta_3 -V(h_0,S_i,\Delta_i)-\Delta T_{h_0}h_0-\sum\limits_{i}\Delta T_{S_i}S_i,
\eeqa
The tadpole cancelation condition is
\beqa
Z_h^{1/2}\Delta T_{h_0}=Z_h^{-1/2}v_{h_0} M^2+\delta \tl{T}_{h_0}=0,
\eeqa
with $\delta \tl{T}$ the one-loop tadpole contributions.
Through the tadpole cancelation condition we can obtain the exact gap equation
\beqa
\mu&=&\f{h_2^2M^2N_c}{8\pi^2M_B^2}\[\sum\limits_{i=1}^3 2\Re(U_{L2i}^{-1*} U_{R1i}^{-1}) \(\la_i-\f{\la_i^3}{M^2}\ln\(\f{M^2+\la_i^2}{\la_i^2}\)\)\f{}{}\],
\eeqa
with the fact that $\mu=Z_h^{-1/2}h_2M v_{h_0}/\sqrt{2}M_B$.
Such form is consistent with the previous large $N_c$ expansion approach with mass insertion.

From the wave function renormalization of the composite Higgs fields, we can get the precise form of the Pagels-Stokar formula
\beqa
v_{h_0}^2&=&\f{\mu^2M^2N_c}{8\pi^2M_B^2}\[\f{}{}\sum\limits_{i=1}^3 2\l|U^{-1*}_{L2i}U_{R1i}^{-1}\r|^2\log\(\f{M^2+\la_i^2}{\la_i^2}\)\.\nn\\
&&+\sum\limits_{i=1}^3\sum\limits_{j=1;i<j}^3(\l| U_{L2i}^{-1*}U_{R1j}^{-1}\r|^2+\l|U_{L2j}^{-1}U_{R1i}^{-1*}\r|^2)\log\(\f{M^2+\la_j^2}{\la_j^2}\)\left.\f{}{}\],
\eeqa
with other possible Higgs VEVs from bottom sector $v_{h_i}^2(i\neq0)$
\beqa
\sum\limits_{i}v_{h_i}^2=v_{EW}^2~.
\eeqa
The VEV of $S_3$ breaks the $U(1)_1$ gauge symmetry completely due to its non-vanishing
$U(1)_1$ quantum number. The expression of $S_3$ will be given in subsequent Section.
Due to the mixing in the Higgs sector, the physical Higgs fields can be obtained by diagonalizing
the relevant mass matrix. We will discuss the complete Higgs sector after we include
the bottom-type quarks.

Similar setting can be seen for the bottom quark. We rewrite the relevant terms for bottom quarks
\beqa
b_L^c &\sim& (1,\bar{3},1)_{(-2/3,0)}~,~\tl{\omega}_L\sim (1,3,1)_{(0,-2/3)}~,~\tl{\sigma}_L\sim (3,1,1)_{(-1/3,-1/3)},\nn\\
\tl{\omega}_L^c&\sim& (\bar{3},1,1)_{(1/3,1/3)}~,~\tl{\sigma}_L^c~\sim (1,\bar{3},1)_{(0,2/3)}~,H_1\sim(1,1,2)_{(-1,0)},~\nn\\H_2&\sim& (1,1,2)_{(1,0)}~,\eeqa
and the induced interactions
\beqa
W\supseteq h\tl{\omega}_L^c \Phi_1\tl{\omega}_L +h\tl{\sigma}_L^c \Phi_2\tl{\sigma}_L+ h \(\bea{c}t_L\\b_L\eea\)H_1\tl{\omega}_L^c+h\(\bea{c}\chi_L^c\\\omega_L^c\eea\)H_2\tl{\sigma}_L
+h\tl{\omega}_L\tl{\sigma}_L^c S_a~.
\eeqa
We can see from the identification that the most general bottom quark mass matrix is
\beqa
(b_L,~\omega_L,~\sigma_L,~\tl{\omega}_L,\tl{\sigma}_L)\(\bea{ccccc}0& s_1& M_1&0 & 0\\\tl{\mu}& M_2&\mu_1&0&\mu_3\\0&0&\mu_2&0&\mu_4\\ \mu_5&0&\mu_6&M_1&\mu_7+s_a\\0&0&0&0&M_2\eea\)\(\bea{c} b_L^c\\ \omega_L^c\\ \sigma_L^c\\\tl{\omega}_L^c\\\tl{\sigma}_L^c\eea\)~.
\eeqa
Similarly, we can diagonalize the mass matrix and obtain the relevant eigenvalues.
We note that the determinant of the mass matrix is
\beqa \det M_{b}=s_1\tl{\mu}\mu_2M_1M_2~, \eeqa which is important to determine
the lightest bottom-type quark masses.
For $M_2= M_1 \gtrsim s_1\gg \mu_i$, we have the eigenvalues of various mass eigenstate in order
\beqa
&&\tl{\la}_2^2 \equiv\tl{m}_{\omega_m}^2\sim M_1^2,~\tl{\la}_3^2\equiv\tl{m}_{\sigma_m}^2\sim 2 M_1^2,~\tl{\la}_4\equiv\tl{m}_{\tl{\omega}_m}^2\sim M_2^2/2,\nn\\
&&\tl{\la}_5^2\equiv \tl{m}_{\tl{\sigma}_m}^2\sim M_2^2/2,
~\tl{\la}_1^2\equiv \tl{m}_{b_m}^2\sim \f{s_1^2\tl{\mu}^2\mu_2^2}{M_1^2M_2^2}~.
\eeqa
Here the expression for the lightest bottom-type quark mass is not precise. This formula is
used to determine the order of the physical bottom mass.
We can also parameterize the mixings in the bottom sector as
\beqa
(b_L^m,~\omega_L^m,~\sigma_L^m,~\tl{\omega}^m_L,\tl{\sigma}^m_L)=Z^L_{ij}(b_L,~\omega_L,~\sigma_L,~\tl{\omega}_L,\tl{\sigma}_L),\nn\\
(b_R^m,~\omega_R^m,~\sigma_R^m,~\tl{\omega}^m_R,\tl{\sigma}^m_R)=Z^R_{ij}(b_R,~\omega_R,~\sigma_R,~\tl{\omega}_R,\tl{\sigma}_R)~.
\eeqa

We can introduce auxiliary fields in symmetry breaking phase to obtain the precise gap equations
\beqa
{\cal L}_\Lambda^b&=&-\tl{\la}_1\bar{b}_L^mb_R^m-\tl{\la}_2\bar{\omega}_L^m\omega_R^m-\tl{\la}_3\bar{\sigma}^m_L\sigma_R^m-\tl{\la}_3\bar{\tl{\omega}}_L^m\tl{\omega}^m_R
-\tl{\la}_5\bar{\tl{\sigma}}_L^m\tl{\sigma}_R^m
-\f{h_2 M}{M_B}\bar{X}_L b_R \tl{H}_1\nn\\
&&-\f{h_2 M}{\sqrt{2}M_B}\bar{X}_L P_R (\Delta_2+S_2+v_{S_2})-\f{h_2 M}{\sqrt{2}M_B}\bar{P}_L P_R (\Delta_3+S_3+v_{S_3})
-\f{h_2 M}{M_B}\bar{\tl{\omega}}_L b_R \tl{S}_4\nn\\
&&-\f{h_2 M}{M_B}\bar{\tl{\omega}}_L P_R \tl{H}_2-\f{h_2 M}{M_B}\bar{X}_L \tl{\sigma}_R \tl{H}_3-\f{h_2 M}{M_B}\bar{P}_L \tl{\sigma}_R \tl{H}_4
-\f{h_2 M}{\sqrt{2}M_B}\bar{\tl{\omega}}_L\tl{\sigma}_R \tl{S}_5\nn\\
&& -\f{h_2 M}{M_B}\bar{\tl{\sigma}}_L \tl{\sigma}_R \tl{S}_6-M^2\(\sum\limits_{i=1}^4|\tl{H}_i|^2+\sum\limits_{i=2}^6|\tl{S}_i|^2+\sum\limits_{i=2}^3|\Delta_i|^2\).
\eeqa
Again we can integrate out the heavy modes and obtain the low energy effective interactions
\beqa
{\cal L}^b_\mu&=&-\tl{\la}_1 \bar{b}_L^mb_R^m-\f{h_2 M}{\sqrt{2}M_B}Z_{L21}^{-1}Z_{R11}^{-1}\bar{b}^m_L b_R^m (h_1+v_{h_1})-\f{h_2 M}{\sqrt{2}M_B}\[Z_{L21}^{-1}Z_{R31}^{-1}\bar{b}^m_L b_R^m (\Delta_{2,0}+S_2+v_{S_2})\.\nn\\
&+&\left.Z_{L31}^{-1}Z_{R31}^{-1}\bar{b}^m_L b_R^m (\Delta_{3,0}+S_3+v_{S_3})\]
-\f{h_2 M}{\sqrt{2}M_B}\left\{\f{}{}Z_{L41}^{-1}Z_{R11}^{-1}\[\bar{b}^m b^m (S_4+v_{S_4})+\bar{b}^m \gamma^5 b^m \tl{\pi}_{S_4}^0\]\right.\nn\\
&+& Z_{L41}^{-1}Z_{R31}^{-1}\[\bar{b}^m b^m (h_2+v_{h_2})+i\bar{b}^m \gamma^5 b^m \pi_{b2}^0\]+Z_{L21}^{-1}Z_{R51}^{-1}\[\bar{b}^m b^m (h_3+v_{h_3})+i\bar{b}^m \gamma^5 b^m \pi_{b3}^0\]\nn\\&+&Z_{L31}^{-1}Z_{R51}^{-1}\[\bar{b}^m b^m (h_4+v_{h_4})+i\bar{b}^m \gamma^5 b^m \pi_{b4}^0\]
+Z_{L41}^{-1}Z_{R51}^{-1}\[\bar{b}^m b^m (S_5+v_{S_5})+\bar{b}^m \gamma^5 b^m \tl{\pi}_{S_5}^0\]\nn\\
&-& \left. Z_{L51}^{-1}Z_{R51}^{-1}\[\bar{b}^m b^m (S_6+v_{S_6})+\bar{b}^m \gamma^5 b^m \tl{\pi}_{S_6}^0\]\f{}{}\right\}
+\f{1}{2}\sum\limits_{i=1}^4Z_{h_i}(\pa_\mu h_i)^2+\f{1}{2}\sum\limits_{i=2}^6
Z_{S_i}(\pa_\mu S_i)^2~\nn\\
&-& Z_{\Delta_i}|\pa_\mu\Delta_i|^2+\f{1}{2}\sum\limits_{i}Z_{S_i}(\pa_\mu S_i)^2-\f{1}{2}Z_{h_0}(\pa_\mu h_0)^2+\f{1}{2}\sum\limits_{i}Z_{S_i}(\pa_\mu S_i)^2+\sum\limits_{i}Z_{\Delta_i}|\pa_\mu\Delta_{i}|^2\nn\\
&-&\f{1}{2}\(\sum\limits_{i=1}^4M_{h_i}^2h_i^2+\sum\limits_{i=2}^6 M_{S_i}^2 S_i^2\)+\sum\limits_{i=2}^3M_{\Delta_i}^2 |\Delta_i|^2
-\sum\limits_{i=2}^6\sum\limits_{j=1}^{4}M_{ij}^{Sh} S_i h_j-\sum\limits_{i=2}^6\sum\limits_{j=2}^{6}M_{ij}^{SS}S_iS_j\nn\\
&-&\sum\limits_{i=1}^4\sum\limits_{j=1}^{4} M_{ij}^{hh}h_ih_j-M_{23}\Delta_2\Delta_3 -V(h_i,S_i,\Delta_i)-\sum\limits_{i=1}^4\Delta T_{h_i}h_i-\sum\limits_{i=2}^6\Delta T_{S_i}S_i,
\eeqa
where we use the parameterization
\beqa \small
\tl{H}_i(i=1,2,3,4)\sim\(\bea{c}\pi_{bi}^+\\ \f{1}{\sqrt{2}}(h_i+\pi_{bi}^0+v_{h_i})\eea\),~\tl{S}_i(i=4,5,6)\sim \f{1}{\sqrt{2}}(S_i+\tl{\pi}_{S_i}^0+v_{S_i})
\eeqa
The tadpole cancelation conditions
\beqa
Z_{h_i}^{1/2}\Delta T_{h_i}=Z_{h_i}^{-1/2}v_{h_i} M^2+\delta \tl{T}_{h_i}=0,\\
Z_{S_i}^{1/2}\Delta T_{S_i}=Z_{S_i}^{-1/2}v_{S_i} M^2+\delta \tl{T}_{S_i}=0,
\eeqa
determine the exact gap equations
\beqa \small
\mu_{H_1}=G_{21}^B,\mu_{H_2}=G_{43}^B,\mu_{H_3}=G_{25}^B,\mu_{H_4}=G_{35}^B,\mu_{S_4}=G_{41}^B,
\mu_{S_5}=G_{45}^B,\mu_{S_6}=G_{55}^B
\eeqa
while $\mu_{S_2}$ and $\mu_{S_3}$ are
\beqa
\mu_{S_2}=G_{23}^B+G_{23}^T~,\mu_{S_3}=G_{23}^B+G_{33}^T~.
\eeqa
Here we use the notation
\beqa
\mu_{H_1}\equiv\tl{\mu},~\mu_{S_2}\equiv \mu_1,~\mu_{S_3}\equiv \mu_2,~\mu_{H_3}{\equiv}\mu_3,~\mu_{H_4}\equiv\mu_4,~\mu_{S_4}\equiv \mu_5,~\mu_{H_2}\equiv\mu_6,~\mu_{S_5}\equiv\mu_7,\nn
\eeqa
and define
\beqa
G^B_{ab}&{\equiv}&\f{h_2^2M^2N_c}{8\pi^2M_B^2}\[\sum\limits_{i=1}^5 2\Re(Z_{Lai}^{-1*} Z_{Rbi}^{-1}) \(\tl{\la}_i-\f{\tl{\la}_i^3}{M^2}\ln\(\f{M^2+\tl{\la}_i^2}{\tl{\la}_i^2}\)\)\]~,\\
G^T_{ab}&{\equiv}&\f{h_2^2M^2N_c}{8\pi^2M_B^2}\[\sum\limits_{i=1}^3 2\Re(U_{Lai}^{-1*} U_{Rbi}^{-1}) \(\la_i-\f{\la_i^3}{M^2}\ln\(\f{M^2+\la_i^2}{\la_i^2}\)\)\].
\eeqa
From the wave function normalization, we can get the Pagels-Stokar formula in the bottom sector
\beqa
v_{h_1}^2&=&\mu_{H_1}^2P_{21}^B~,~v_{h_2}^2=\mu_{H_2}^2P_{43}^B~,~v_{h_3}^2=\mu_{H_3}^2P_{25}^B~,~v_{h_4}^2=\mu_{H_4}^2P_{35}^B~,
\eeqa
with the relation $\sum\limits_{i=0}^4v_{h_i}^2=v_{EW}^2$ as well as the Pagels-Stokar formula for $S_i$
\beqa
&&v_{S_2}^2 =\mu_{S_2}^2\(P_{23}^B+P_{23}^T\),
~v_{S_3}^2 =\mu_{S_3}^2\(P_{33}^B+P_{33}^T\),\nn\\
&&
v_{S_4}^2 =\mu_{S_4}^2 P_{41}^B,~v_{S_5}^2 =\mu_{S_5}^2 P_{45}^B, ~v_{S_6}^2 =\mu_{S_6}^2 P_{55}^B,
\eeqa
with $v_{S_3}^2$ the $U(1)_2$ breaking scale and the fact $\mu_{S_3}=Z_{S_3}^{-1/2}h_2M v_{S_3}/\sqrt{2}M_B$.
Here, we define
\beqa\tiny
P_{ab}^B&\equiv&\f{M^2N_c}{8\pi^2M_B^2}\[\f{}{}\sum\limits_{i=1}^5 2\l|Z^{-1*}_{Lai}Z_{Rbi}^{-1}\r|^2\log\(\f{M^2+\tl{\la}_i^2}{\tl{\la}_i^2}\)\.\nn\\
&&+\sum\limits_{i=1}^5\sum\limits_{i,j=1;i<j}^5(\l| Z_{Lai}^{-1*}Z_{Rbj}^{-1}\r|^2+\l|Z_{Laj}^{-1}Z_{Rbi}^{-1*}\r|^2)\log\(\f{M^2+\tl{\la}_{j}^2}{\tl{\la}_{j}^2}\)
\left.\f{}{}\]~,\nn\\
P_{ab}^T&\equiv&\f{M^2N_c}{8\pi^2M_B^2}\[\f{}{}\sum\limits_{j=1}^3 2\l|U^{-1*}_{Lai}U_{Rbi}^{-1}\r|^2\log\(\f{M^2+\la_j^2}{\la_j^2}\)\.\nn\\
&&+\sum\limits_{j=1}^3\sum\limits_{k=1;j<k}^3 (\l| U_{Lai}^{-1*}U_{Rbj}^{-1}\r|^2+\l|U_{Laj}^{-1}U_{Rbi}^{-1*}\r|^2)\log\(\f{M^2+\la_j^2}{\la_j^2}\)\left.\f{}{}\]~.
\eeqa
\normalsize
The physical Higgs fields can be obtained by diagonalizing the $10\times 10$ mixing mass matrix between $h_i (i=0,\cdots,4)$ and $S_j(j=2,\cdots,6)$.
Each entry can be calculated by the one-loop diagrams in the large $N_c$ fermion bubble approximation. Detailed expressions can be found in appendix B.

One combination of $\pi_t^{0,\pm}$ and $\pi_{bi}^{0,\pm}$ will act as ``would be'' Goldstone bosons
which will be eaten by $W^\pm$ and $Z_0$. The remaining $\pi_{t,bi}^{\pm}$ will combine into multiple
charged Higgs fields $H^{\pm}_i$ while the other combinations of $\pi_{bi}^{0}$ and $\pi_t^0$ will be
the CP-odd Higgs fields $A_i^0$.
The mixings between triplet Higgs fields will give two mass eigenstates $\tl{\Delta}_{2}$ and
$\tl{\Delta}_{3}$.
There is enough parameter space to tune the lightest Higgs field to be at 125 GeV.
We note that the non-minimal nature is crucial for Higgs mixing and the appearance of light Higgs field.

Quarks of the first two generations transform as $SU(3)_2$ fundamental representations and also
carry $U(1)_1$ charges. As $SU(3)_2$ will become strongly coupled, additional $U(1)_1$ interactions
can prevent the condensation between the first two generations. This is similar to that of the
flavor-universal topcolor model~\cite{universalcoloron}.

The most important electroweak precision constraints on top seesaw comes from the
electroweak oblique parameters $S$ and $T$~\cite{STU}.
Minimal Top seesaw model can non-trivially satisfy the $S-T$ bounds. We know that the
oblique parameter $S$ can be thought of as the measure of the total size of the new sector
while $T$ is the measure of the weak-isospin breaking induced by it. Just as ordinary
extended Top Seesaw model with bottom seesaw, the contributions to the oblique parameters are rather complicate. Detailed analytic expressions for new contributions to $S,T$ parameters can be seen in appendix C. Although the precise values need the detailed numerical studies, we note that the contributions
to the $S$ parameter should be very similar to that of the minimal top seesaw model
because most new particle contents are vector-like.

The contributions from the multiple Higgs doublets needs the Higgs spectrum as well as
the knowledge of the mixings among different Higgs doublets. In general, they should drive
the $T$ parameter negative which however being compensated by isospin violating quark sector
contributions. We will left the detailed numerical results to subsequence studies.
We just anticipate that there are enough parameter space to make our model compatible with $S-T$ bounds.

There are additional constraints from $Z-b_L-b_L$ coupling. The mixing within the bottom seesaw
sector change the vertex by
\beqa
\delta g_L^b=\f{e}{2\sin\theta\cos\theta}\(\sum\limits_{i=4}^5 |Z_{1j}|^2\)~.
\eeqa
We can see that $\Gamma(Z\ra \bar{b}b)$ will decrease with respect to the SM predictions.
The updated data on $R_b$ will constrain the mixings within the bottom sector.

We can properly choose the parameter $M_1=M_2=20 {\rm ~TeV}$ so that the physical top quark mass is given by
\beqa
\la_1^2\approx\f{\mu^2\mu_2^2s_1^2}{M_1^2M_2^2}\approx(170{\rm GeV})^2~.
\eeqa
The gap equation depends implicitly on $\mu$ and $\mu_2$ on the r.h.s and
we checked that the following parameters
\beqa
s_1\simeq 18 {\rm ~TeV} , \mu_2\simeq 5.02 {\rm ~TeV}, \mu\simeq 0.76{\rm ~TeV}
\eeqa
can satisfy approximately the gap equation
\beqa
\f{\mu}{\mu_2}\approx \f{U_{L33}^{-1*}U_{R33}^{-1}}{U_{L23}^{-1}U_{R13}^{-1}}.
\eeqa
The mixing matrices can be obtained by diagonalizing the mass matrices
\beqa
\(\bea{c}
t_L^m\\
\chi_L^m\\
\rho_L^m
\eea\)&=&\(\bea{ccc}
-0.2475&0.4940&-0.8335\\
0.2225&-0.8083&-0.5451\\
0.9430&0.3204&-0.0902
\eea\)
\(\bea{c}
t_L\\
\chi_L\\
\rho_L
\eea\)~,\\
\(\bea{c}
t_R^m\\
\chi_R^m\\
\rho_R^m
\eea\)&=&\(\bea{ccc}
0.9988&-0.0475&-0.0132\\
-0.03766&-0.5623&-0.8261\\
0.03180&0.8256&-0.5634
\eea\)\(\bea{c}
t_R\\
\chi_R\\
\rho_R
\eea\),
\eeqa
with the eigenvalues
\beqa
\la_1\simeq 0.172{\rm ~TeV}, \la_2\simeq 13{\rm ~TeV}~, \la_3\simeq 31.36{\rm ~TeV}~.
\eeqa
The unitary nature of the mixing matrix indicates that the inverse mixing matrix is the form
\beqa
U_L^{-1}=\(\bea{ccc}
-0.2475&0.2225&0.9430\\
0.4940&-0.8083&0.3204\\
-0.8335&-0.5451&-0.0902
\eea\),
U_R^{-1}=\(\bea{ccc}
0.9988&-0.03766&0.03180\\
-0.0475&-0.5623&0.8256\\
-0.0132&-0.8261&-0.5634
\eea\)
\eeqa
From the Pagels-Stokar formula and setting $v_t^2\simeq (200 {\rm GeV})^2$ and $N_c=3$,
we obtain the cut-off scale
\beqa
M\simeq 40~{\rm TeV}~.
\eeqa
The coloron mass scale is $M_B\simeq \sqrt{2}M_1\approx 30 ~{\rm TeV}$ if we assume $h_2 \sim {\cal O}(1)$.
The bottom quark sector can be similarly obtained. The lightest bottom-type quark mass is given approximately by
\beqa
\tl{\la}_1\approx\f{\tl{\mu}s_1 \mu_2}{M_1M_2}=4.2~{\rm GeV}~,
\eeqa
which is related to the top quark sector through the relation
${\mu}/{\tl{\mu}}{\equiv}\tan\beta_1\approx 40$.

As indicated in section \ref{sec-1}, most superpartners obtain their masses via gauge mediation. For proper chosen $M$ with $\sqrt{F_M}/M\sim {\cal O}(1)$, the dominant gauge mediated contributions to sfermion masses come from the nearly strong $U(1)_1$ and $SU(3)_2$ gauge interactions. Then from the gauge mediated supersymmetry breaking formula, we can easily set the soft mass parameters to lie near $\sqrt{F}\sim 20~{\rm TeV}$. Thus, below $\sqrt{F}$ after we integrate out the sfermion fields, the low energy effective theory reduce to NJL type top seesaw interactions.

While the superpartners are lighter than the coloron, their contribution to the four-fermion interactions are subdominant because of the R-parity. Possible four-fermion interactions contributed from superpartners in the low energy can be only generated by sparticle loops which thus amount to the suppression scale of the operators to be $4\pi\times 20{\rm TeV} \sim 200 {\rm TeV}$.

  In general, the scalar type bound state of the NJL-type condensation has
a mass of order $2\mu$ with $\mu$ the corresponding dynamical mass in the gap equation.
In our scenario, the lightest scalar states are mixing between various condensation bound states
with the lightest bound state as light as $2\tl{\mu}\sim {\cal O}(10~{\rm GeV})$, and
then can be as light as ${\cal O}(100~{\rm GeV})$.

 We should note that quite a bit fine tuning is necessary in our scenario. By introducing the auxiliary fields, dynamical Higgs field will reappear after renormalization group equation running down to a lower scale. Thus fine tuning problem that is plaguing the ordinary Higgs models will also show up here as long as the cut off scale is not too low. In our scenario, the cut off scale of the NJL type interaction is 40 TeV, thus leads to fine tuning of order
\beqa
\(\f{\Lambda}{4\pi m_h}\)^2\sim 600.
\eeqa
As there are much parameter space remaining in our scenario, it may be possible to ameliorate such fine tuning by other choices of parameters. We leave the detailed numerical discussions in our subsequent papers.

 We would like to give a brief comment on the status of this model in the LHC era. In the previous choice of the parameter, new fermions will acquire masses of order $M_1^2$ and $M_2^2$ [$\sim {\cal O}(10 {\rm TeV})$] thus cannot be discovered by LHC. In the low energy, our theory will look like a two Higgs-doublet model with the mixings between top-Higgs and bottom-higgs to give the 125 GeV scalar that was discovered by LHC. The light scalar in our scenario is standard model-like with its couplings to $W,Z$ gauge bosons and photons resembling that of the standard model Higgs field. While the detailed mass paramters of the additional Higgs fields depending on the concrete values of the mixing and Renormalization Group Equation running, a coarse estimation on the tree-level mass of the CP-odd Higgs field is that $M_{A^0}\approx 350 {\rm GeV}$; the charged Higgs $H^\pm$ have masses $M_{H^\pm}\approx \sqrt{M_A^2+m_W^2}\sim 359{\rm GeV}$; the heavy CP-even higgs $H^0$ acquires a mass $m_{H^0}\sim M_{A^0}$. Our predictions on the Higgs masses are not very sensitive to the UV physics, so it is testable on the LHC. Inspired by this work, a phenomenological low energy top-bottom seesaw model which can explain the LHC discoveries is being studied in our new paper.

\section{\label{sec-3}Conclusions}

The recent discovery of a 125~GeV Higgs-like particle at the LHC pushes us to ask the interesting
question whether such scalar is composite or fundamental. On the other hand, top quark, which is much
heavier than all the other SM fermions, indicates that it couples more strongly to electroweak
symmetry breaking sector. Thus, it is possible that the top sector plays a key role in electroweak
symmetry breaking mechanism and related intimately to the intrinsic nature of the Higgs field.
Ordinary top seesaw model predicts too heavy Higgs mass and requires new matter contents and
interactions that are put in by hand. We propose a typical non-minimal extended top seesaw model
(with also bottom seesaw) and accommodate a light composite Higgs field. The non-minimal nature
is crucial for Higgs mixing and the appearance of light Higgs field. Besides, supersymmetric
strong dynamics can lead to almost composite top and bottom quark as well as new emergent
topcolor gauge interactions. At the same time, supersymmetry breaking condition also leads to
topcolor breaking as well. The low energy QCD coupling is partially emergent. This theory also
acts as an AdS/CFT dual to a Randall-Sundrum~\cite{rs} type model which will be given in subsequent studies.

\begin{acknowledgments}
We are very grateful to the referee for enlightenment discussions and comments.
Fei Wang and Csaba Balazs acknowledges Kavli Institute of Theoretical Physics China (KITPC) for its excellent research
environments during their staying in Beijing.
This research was supported by the Australian Research Council under project DP0877916,
by the Natural Science Foundation
of China under grant numbers 11105124, 11075003, 10821504, 11075194, and 11135003,
by the Project of Knowledge Innovation Program (PKIP) of Chinese Academy of Sciences
under grant number KJCX2.YW.W10,
and by the DOE grant DE-FG03-95-Er-40917.

\section{Appendix A: Gap Equation}
There are several ways to obtain the mass gap of the dynamical condensations. The condensations can be calculated from the gap equations which are large-$N_c$ Dyson-Schwinger equations expanded up to ${\cal O}(m_{\chi t}^3)$ for the NJL Lagrangian. The relevant diagrams are shown in fig(1).
\begin{figure}[tb]
\label{gap}
\centering
\includegraphics[width=10cm]{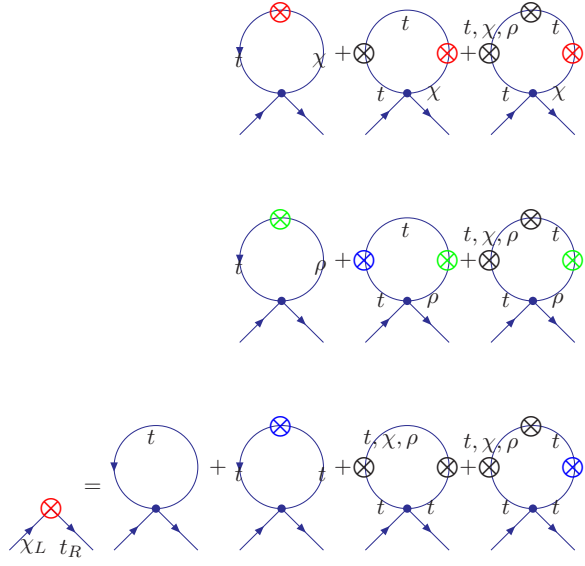}
\caption{The gap equations for quark condensations in the top sector. The red, blue and green crosses
denote the $\bar{t}^\pr\chi^\pr$, $\bar{t}^\pr{t}^\pr$ and $\bar{t}^\pr\rho^\pr$ condensations, respectively.
While black crosses denote all the previous three condensations. } \label{top}
\end{figure}
 Tedious calculations give the expression for $\bar{\chi}_L^\pr t_R^\pr$ condensation
\beqa \small
\mu N_{22}^{-1}&\approx&\f{h^2N_c}{4\pi^2 M_B^2} \left\{ N_{22}^3 \mu\[\f{}{}M^2-\(\ol{M}_1^2+N_{21}\mu \ol{M}_1\)\ln\(\f{\ol{M}_1+M^2}{\ol{M}_1^2}\)
\right.\nn\.\\&-&(N_{21}^2+N_{22}^2+N_{23}^2)\mu^2\log\(\f{M^2+\overline{M}_1^2}{\overline{M}_1^2}\)
+\left.\f{N_{23}^2\mu^2M_1^2}{M_2^2-M_1^2}\log\(\f{(M^2+M_1^2)M_2^2}{(M^2+M_2^2)M_1^2}\) \]~\nn\\
&+& N_{22} N_{23}^2\mu\[\f{}{}M^2-\overline{M}_2^2\log\(\f{M^2+\overline{M}_2^2}{\overline{M}_2^2}\)
-\(\ol{M}_2^2+N_{21}\mu\ol{M}_2\)\log\(\f{M^2+\overline{M}_2^2}{\overline{M}_2^2}\)\.\nn\\
&-&\left.\(N_{21}^2+N_{22}^2+N_{23}^2\)\mu^2\log\(\f{M^2+\overline{M}_2^2}{\overline{M}_2^2}\)
+ \f{N_{21}^2\mu^2 M_2^2}{(M_1^2-M_2^2)}\log\(\f{(M^2+M_2^2)M_1^2}{(M^2+M_1^2)M_2^2}\) \]\nn\\
&+& N_{22} N_{21}^2\mu\[\f{}{}M^2
-\f{N_{22}^2}{N_{21}}\mu \ol{M}_1\log\(\f{M^2+\overline{M}_1^2}{\overline{M}_1^2}\)-\f{N_{23}^2}{N_{21}}\mu \ol{M}_2\log\(\f{M^2+\overline{M}_2^2}{\overline{M}_2^2}\)\.\nn\\
&-&\left.\left.\(N_{21}^2+N_{22}^2+N_{23}^2\)\mu^2\log\(\f{M^2+\overline{M}_2^2}{\overline{M}_2^2}\)+ N_{22}^2\mu^2 M_2^2\log\(\f{(M^2+M_1^2)M_2^2}{(M^2+M_2^2)M_1^2}\) \]\right\}.\nn\\
\eeqa
From this expression, we can easily deduce the analytic expression for the form of effective critical coupling.
Similarly, we can get the other condensations
\beqa
<\bar{t}_L^\pr \rho_R^\pr>,<\bar{\chi}_L^\pr \rho_R^\pr>,<\bar{\rho}^\pr \rho_R^\pr>, \cdots
\eeqa
to give $<\bar{X}_L P_R>$ and $<\bar{P}_L P_R>$.
\section{Appendix B: Mixing In the Higgs Sector}
The CP-even Higgs fields in our scenario are obtained by diagonalize the $10\times 10$ mass matrix. In the fermion bubble approximation, the diagonal entry can be calculated to be
\beqa
&&m_{h_0}^2=\f{M_{21}^T}{Z_{h_0}}~,~m_{h_1}^2=\f{M_{21}^B}{Z_{h_1}}~,~m_{h_2}^2=\f{M_{43}^B}{Z_{h_2}}~,~m_{h_3}^2=\f{M_{25}^B}{Z_{h_3}}~,~m_{h_4}^2=\f{M_{35}^B}{Z_{h_4}},\nn\\
&&m_{S_2}^2=\f{M_{23}^T+M_{23}^B-M^2}{Z_{S_2}},~m_{S_3}^2=\f{M_{33}^T+M_{33}^B-M^2}{Z_{S_3}}, \nn\\
&& m_{S_4}^2=\f{M_{41}^B}{Z_{S_4}}~,m_{S_5}^2=\f{M_{45}^B}{Z_{S_5}}~,
m_{S_6}^2=\f{M_{55}^B}{Z_{S_6}}~,
\eeqa
where the wave function renormalizations are
\beqa
&&Z_{h_0}=\f{h_2^2}{2}P_{21}^T,~Z_{h_1}=\f{h_2^2}{2}P_{21}^B,~Z_{h_2}=\f{h_2^2}{2}P_{43}^B,~Z_{h_3}=\f{h^2}{2}P_{25}^B,~Z_{h_4}=\f{h_2^2}{2}P_{35}^B,,\nn\\
&&Z_{S_2}=\f{h_2^2}{2}\(P_{23}^B+P_{23}^T\),
~Z_{S_3}=\f{h_2^2}{2}\(P_{33}^B+P_{33}^T\),\nn\\
&&Z_{S_4}=\f{h_2^2}{2}P_{41}^B~,~Z_{S_5}=\f{h_2^2}{2}P_{45}^B~,
~Z_{S_6}=\f{h_2^2}{2}P_{55}^B~,
\eeqa
and the definitions for $M_{ab}^T$ and $M_{ab}^B$ are
\beqa
M_{ab}^T&\equiv&\left\{(1-\f{h_2^2M^2N_c}{8\pi^2M_B^2})M^2+\f{h_2^2M^2N_c}{8\pi^2M_B^2}\[4\sum\limits_{i=1}^3\(\Re (U_{Lai}^{-1*}U_{Rbi}^{-1})\)^2\la_i^2\ln\(\f{M^2}{\la_i^2}\)\.\.\nn\\
&+&\sum\limits_{i,j=1;i>j}^32\Re[(U_{Lai}^{-1*}U_{Rbj}^{-1})(U_{Laj}^{-1*}U_{Rbi}^{-1})]\f{\la_i\la_j}{(\la_i^2-\la_j^2)}[\la_i^2\ln\f{M^2}{\la_i^2}-\la_j^2\ln\f{M^2}{\la_j^2}]\nn\\
&+&\sum\limits_{i,j=1;i>j}^3\[\l|U_{Lai}^{-1*}U_{Rbj}^{-1}\r|^2+\l|U_{Laj}^{-1*} U_{Rbi}^{-1}\r|^2\]\f{1}{\la_i^2-\la_j^2}[\la_i^4\ln\f{M^2}{\la_i^2}-\la_j^4\ln\f{M^2}{\la_j^2}]\left.\f{}{}\]\left.\f{}{}\right\},\nn\\
M_{ab}^B&\equiv&\left\{(1-\f{h_2^2M^2N_c}{8\pi^2M_B^2})M^2+\f{h_2^2M^2N_c}{8\pi^2M_B^2}
\[4\sum\limits_{i=1}^3\(\Re (Z_{Lai}^{-1*}Z_{Rbi}^{-1})\)^2\tl{\la}_i^2\ln\(\f{M^2}{\tl{\la}_i^2}\)\.\.\nn\\
&+&\sum\limits_{i,j=1;i>j}^52\Re[(Z_{Lai}^{-1*}Z_{Rbj}^{-1})(Z_{Laj}^{-1*}Z_{Rbi}^{-1})]\f{\tl{\la}_i\tl{\la}_j}{(\tl{\la}_i^2-\tl{\la}_j^2}[\tl{\la}_i^2\ln\f{M^2}{\tl{\la}_i^2}-\tl{\la}_j^2\ln\f{M^2}{\tl{\la}_j^2}] \nn\\
&+&\sum\limits_{i,j=1;i>j}^5\[\l|Z_{Lai}^{-1*}Z_{Rbj}^{-1}\r|^2+\l|Z_{Laj}^{-1*} Z_{Rbi}^{-1}\r|^2\] \f{1}{\tl{\la}_i^2-\tl{\la}_j^2}[\tl{\la}_i^4\ln\f{M^2}{\tl{\la}_i^2}-\tl{\la}_j^4\ln\f{M^2}{\tl{\la}_j^2}]\left.\f{}{}\]\left.\f{}{}\right\}~.\nn\\
\eeqa

The mixings between the Higgs fields can be calculated accordingly.
For simplicity, we can define
\beqa
F_{ab,cd}^{BB}&\equiv&
-\f{h_2^2M^2N_c}{8\pi^2M_B^2}M^2
+\f{h_2^2M^2N_c}{8\pi^2M_B^2}\[\sum\limits_{i=1}^54 \Re\(Z_{Lai}^{-1*}Z_{Rbi}^{-1}\)\Re\(Z_{Lci}^{-1*}Z_{Rdi}^{-1}\)\tl{\la}_i^2\ln\(\f{M^2}{\tl{\la}_i^2}\)\.\nn\\
&&+\sum\limits_{i,j=1;i>j}^5[(Z_{Lai}^{-1*}Z_{Rbj}^{-1})(Z_{Lcj}^{-1*}Z_{Rdi}^{-1})
+(Z_{Laj}^{-1}Z_{Rbi}^{-1*})(Z_{Lci}^{-1}Z_{Rdj}^{-1*})] \nn\\
&& \times \f{\tl{\la}_i\tl{\la}_j}{(\tl{\la}_i^2-\tl{\la}_j^2)}[\tl{\la}_i^2\ln\f{M^2}{\tl{\la}_i^2}-\tl{\la}_j^2\ln\f{M^2}{\tl{\la}_j^2}]
+\sum\limits_{i,j=1;i>j}^5\[(Z_{Lai}^{-1*}Z_{Rbj}^{-1})(Z_{Lci}^{-1} Z_{Rdj}^{-1*})
\nn\.\\
&&\left.
+(Z_{Laj}^{-1}Z_{Rbi}^{-1*})(Z_{Lcj}^{-1*}Z_{Rdi}^{-1})\]\f{1}{\tl{\la}_i^2-\tl{\la}_j^2}[\tl{\la}_i^4\ln\f{M^2}{\tl{\la}_i^2}-\tl{\la}_j^4\ln\f{M^2}{\tl{\la}_j^2}]\left.\f{}{}\], \\
F_{ab,cd}^{TT}&\equiv&
-\f{h_2^2M^2N_c}{8\pi^2M_B^2}M^2+\f{h_2^2M^2N_c}{8\pi^2M_B^2}\[\sum\limits_{i=1}^3 4\Re\(U_{Lai}^{-1*}U_{Rbi}^{-1}\)\Re\(U_{Lci}^{-1*}U_{Rdi}^{-1}\){\la}_i^2\ln\(\f{M^2}{{\la}_i^2}\)\.\nn\\
&&+\sum\limits_{i,j=1;i>j}^3[(U_{Lai}^{-1*}U_{Rbj}^{-1})(U_{Lcj}^{-1*}U_{Rdi}^{-1})
+(U_{Laj}^{-1}U_{Rbi}^{-1*})(U_{Lci}^{-1}U_{Rdj}^{-1*})]\nn\\
&&\times
\f{{\la}_i{\la}_j}{({\la}_i^2-{\la}_j^2)}[{\la}_i^2\ln\f{M^2}{{\la}_i^2}-{\la}_j^2\ln\f{M^2}{{\la}_j^2}]
+\sum\limits_{i,j=1;i>j}^3\f{1}{{\la}_i^2-{\la}_j^2}\[(U_{Lai}^{-1*}U_{Rbj}^{-1})\right.\nn\\
&&\left.\times(U_{Lci}^{-1} U_{Rdj}^{-1*})
+(U_{Laj}^{-1}U_{Rbi}^{-1*})(U_{Lcj}^{-1*}U_{Rdi}^{-1})\][{\la}_i^4\ln\f{M^2}{{\la}_i^2}-{\la}_j^4\ln\f{M^2}{{\la}_j^2}]
\left.\f{}{}\],
\eeqa
where the mixing terms between $h_i$ and $h_j$ are
\beqa
&&M_{12}^{hh}=\f{F_{21,43}^{BB}}{\sqrt{Z_{h_1}Z_{h_2}}}~,M_{13}^{hh}=\f{F_{21,25}^{BB}}{\sqrt{Z_{h_1}Z_{h_3}}}~,M_{14}^{hh}=\f{F_{21,35}^{BB}}{\sqrt{Z_{h_1}Z_{h_4}}}~,\nn\\
&& M_{23}^{hh}=\f{F_{43,25}^{BB}}{\sqrt{Z_{h_2}Z_{h_3}}}~
M_{24}^{hh}=\f{F_{43,35}^{BB}}{\sqrt{Z_{h_2}Z_{h_4}}}~,M_{34}^{hh}=\f{F_{25,35}^{BB}}{\sqrt{Z_{h_3}Z_{h_4}}},~M_{0i}^{hh}=0~,
\eeqa
the mixing terms within $S_i$ are
\beqa
&&M_{23}^{SS}=\f{F_{23,33}^{TT}+F_{23,33}^{BB}}{\sqrt{Z_{S_2}Z_{S_3}}}~,
M_{24}^{SS}=\f{F_{23,41}^{BB}}{\sqrt{Z_{S_2}Z_{S_4}}}~,
M_{25}^{SS}=\f{F_{23,45}^{BB}}{\sqrt{Z_{S_2}Z_{S_5}}}~,\nn\\
&&M_{26}^{SS}=\f{F_{23,55}^{BB}}{\sqrt{Z_{S_2}Z_{S_6}}}~
M_{34}^{SS}=\f{F_{33,41}^{BB}}{\sqrt{Z_{S_3}Z_{S_4}}}~,
M_{35}^{SS}=\f{F_{33,45}^{BB}}{\sqrt{Z_{S_3}Z_{S_5}}}~, \nn\\
&& M_{36}^{SS}=\f{F_{33,55}^{BB}}{\sqrt{Z_{S_3}Z_{S_6}}}~,
M_{45}^{SS}=\f{F_{41,45}^{BB}}{\sqrt{Z_{S_4}Z_{S_5}}}~,
M_{46}^{SS}=\f{F_{41,55}^{BB}}{\sqrt{Z_{S_4}Z_{S_6}}}~,
\eeqa
the $h_i$ and $S_j$ mixing terms are
\beqa
M_{02}^{hS}&=&\f{F_{21,23}^{TT}}{\sqrt{Z_{h_0}Z_{S_2}}}~,M_{03}^{hS}=\f{F_{21,33}^{TT}}{\sqrt{Z_{h_0}Z_{S_3}}}~,M_{04}^{hS}=M_{05}=M_{06}=0~,
M_{12}^{hS}=\f{F_{21,23}^{BB}}{\sqrt{Z_{h_1}Z_{S_2}}}~,\nn\\ M_{13}^{hS}&=&\f{F_{21,33}^{BB}}{\sqrt{Z_{h_1}Z_{S_3}}}~,M_{14}^{hS}=\f{F_{21,41}^{BB}}{\sqrt{Z_{h_1}Z_{S_4}}}~,
M_{15}^{hS}=\f{F_{21,45}^{BB}}{\sqrt{Z_{h_1}Z_{S_5}}}~,M_{16}^{hS}=\f{F_{21,55}^{BB}}{\sqrt{Z_{h_1}Z_{S_6}}}~,\nn\\
M_{22}^{hS}&=&\f{F_{43,23}^{BB}}{\sqrt{Z_{h_2}Z_{S_2}}}~,M_{23}^{hS}=\f{F_{43,33}^{BB}}{\sqrt{Z_{h_2}Z_{S_3}}}~,M_{24}^{hS}=\f{F_{43,41}^{BB}}{\sqrt{Z_{h_2}Z_{S_4}}}~,
M_{25}^{hS}=\f{F_{43,45}^{BB}}{\sqrt{Z_{h_2}Z_{S_5}}}~,\nn\\
M_{26}^{hS}&=&\f{F_{43,55}^{BB}}{\sqrt{Z_{h_2}Z_{S_6}}}~,M_{32}^{hS}=\f{F_{25,23}^{BB}}{\sqrt{Z_{h_3}Z_{S_2}}}~,M_{33}^{hS}=\f{F_{25,33}^{BB}}{\sqrt{Z_{h_3}Z_{S_3}}}~,
M_{34}^{hS}=\f{F_{25,41}^{BB}}{\sqrt{Z_{h_3}Z_{S_4}}}~,\nn\\ M_{35}^{hS}&=&\f{F_{25,45}^{BB}}{\sqrt{Z_{h_3}Z_{S_5}}}~,M_{36}^{hS}=\f{F_{25,55}^{BB}}{\sqrt{Z_{h_3}Z_{S_6}}}~,
M_{42}^{hS}=\f{F_{35,23}^{BB}}{\sqrt{Z_{h_4}Z_{S_2}}}~,
M_{43}^{hS}=\f{F_{35,33}^{BB}}{\sqrt{Z_{h_4}Z_{S_3}}}~,\nn\\
M_{44}^{hS}&=&\f{F_{35,41}^{BB}}{\sqrt{Z_{h_4}Z_{S_4}}}~,M_{45}^{hS}=\f{F_{35,45}^{BB}}{\sqrt{Z_{h_4}Z_{S_5}}}~,
M_{46}^{hS}=\f{F_{35,55}^{BB}}{\sqrt{Z_{h_4}Z_{S_6}}}~,
\eeqa
and the relations are
\beqa
M_{ij}^{hS}&=&\f{F_{ab,cd}^{TT,BB}}{\sqrt{Z_{h_i}Z_{S_j}}}~,~
M_{ji}^{hS}=\f{F_{cd,ab}^{TT,BB}}{\sqrt{Z_{S_i}Z_{h_j}}}~.
\eeqa
\section{Appendix C: Oblique Parameters}
 The most important constraints of out model is the oblique parameters. New contributions to the $T$ parameter
\beqa
T=\f{4\pi}{\sin^2\theta\cos^2\theta M_Z^2}\[\Pi_{11}\l|_{q^2=0}-\Pi_{33}\r|_{q^2=0}\]~,
\eeqa
from the quark sector are
\beqa
\delta T&=&\f{4\pi}{s_W^2c_W^2 m_Z^2}\f{4 N_c}{16\pi^2}\f{1}{4}\left\{\f{}{} \sum\limits_{a=1}^3\sum\limits_{b=1}^5 \[\l|\sum\limits_{i=1}^3U_{Lia}^{-1*}Z_{Lib}^{-1}\r|^2 +\l|\sum\limits_{i=2}^3U_{Ria}^{-1*} Z_{Rib}^{-1}\r|^2 \] K(\la_a,\tl{\la}_b)\.\nn\\
&+&\sum\limits_{a=1}^3\sum\limits_{b=1}^5 2\Re \[\(\sum\limits_{i=1}^3U_{Lia}^{-1*}Z_{Lib}^{-1}\)\(\sum\limits_{i=2}^3U_{Ria}^{-1} Z_{Rib}^{-1*}\)\]L(\la_a,\tl{\la}_b)\nn\\
&-& \sum\limits_{a=1}^3 \l| \(\sum\limits_{i=1}^3 U_{Lia}^{-1*}U_{Lia}^{-1}\)+\(\sum\limits_{i=2}^3U_{Ria}^{-1*}U_{Ria}^{-1}\)\r|^2 \la_a^2\log \f{\la_a^2}{M_B^2}\nn\\
&-&\sum\limits_{a=1}^5\l|\(\sum\limits_{i=1}^3 Z_{Lia}^{-1*}Z_{Lia}^{-1}\)+\(\sum\limits_{i=2}^3 Z_{Ria}^{-1*}Z_{Ria}^{-1}\)\r|^2 \tl{\la}_a^2\log\f{\tl{\la}_a^2}{M_B^2}\nn\\
&-&\sum\limits_{a=1}^3\sum\limits_{b=1;b\neq a}^3 \[\left|\sum\limits_{i=1}^3 U_{Lia}^{-1*}U_{Lib}^{-1}\right|^2
+\left|\sum\limits_{i=2}^3 U_{Ria}^{-1*} U_{Rib}^{-1}\right|^2 \] K(\la_a,{\la}_b)\nn\\
&-&\sum\limits_{a=1}^3\sum\limits_{b=1;b\neq a}^3 2 \Re \[\(\sum\limits_{i=1}^3(U_{Lia}^{-1*}U_{Lib}^{-1}\)\(\sum\limits_{i=2}^3 U_{Ria}^{-1}U_{Rib}^{-1*}\)\]L(\la_a,{\la}_b)\nn\\
&-&\sum\limits_{a=1}^5\sum\limits_{b=1;a\neq b}^5 \[\l|\sum\limits_{i=1}^3Z_{Lia}^{-1*}Z_{Lib}^{-1}\r|^2 +\l|\sum\limits_{i=2}^3Z_{Ria}^{-1*} Z_{Rib}^{-1}\r|^2 \] K(\tl{\la}_a,\tl{\la}_b)\nn\\
&-&\sum\limits_{a=1}^5\sum\limits_{b=1;a\neq b}^5 2\Re \[ \(\sum\limits_{i=1}^3Z_{Lia}^{-1*}Z_{Lib}^{-1}\)\(\sum\limits_{i=2}^3Z_{Ria}^{-1}Z_{Rib}^{-1*}\)\]L(\tl{\la}_a,\tl{\la}_b)
-\f{1}{4}\la_1^2\left.\f{}{}\right\},
\eeqa
here we define
\beqa
K(a,b)&\equiv &\f{1}{a^2-b^2}\[\f{a^4}{2}\ln\f{a^2}{M_B^2}-\f{b^4}{2}\ln\f{b^2}{M_B^2}-\f{1}{4}a^4+\f{1}{4}b^4\]~,\nn\\
L(a,b)&\equiv &\f{ab}{a^2-b^2}\[a^2\ln\f{a^2}{M_B^2}-b^2\ln\f{b^2}{M_B^2}-a^2+b^2\]~.
\eeqa
New contributions to the oblique $S$ parameter
\beqa
S={16\pi}\[\f{\pa}{\pa q^2}\Pi_{33}\l|_{q^2=0}-\f{\pa}{\pa q^2}\Pi_{3Q}\r|_{q^2=0}\]~,
\eeqa
from the quark sector are
\scriptsize
\beqa
-\f{4\pi S}{N_c}&=&
\f{1}{3}\sum\limits_{a=1}^3\[\(\sum\limits_{i=1}^3 U_{Lia}^{-1*}U_{Lia}^{-1}- \sum\limits_{i=2}^3U_{Ria}^{-1*}U_{Ria}^{-1}-4U_{R1a}^{-1*}U_{R1a}\)
\(\sum\limits_{i=1}^3 U_{Lia}^{-1*}U_{Lia}^{-1}-\sum\limits_{i=2}^3 U_{Ria}^{-1*}U_{Ria}^{-1}\)\] \nn\\
&+&\f{2}{9}\sum\limits_{a=1}^3\ln(\la_a^2)\[ \(\sum\limits_{i=1}^3 U_{Lia}^{-1*}U_{Lia}^{-1}\)\(\sum\limits_{i=1}^3 U_{Lia}^{-1*}U_{Lia}^{-1}\)+\(\sum\limits_{i=2}^3 U_{Ria}^{-1*}U_{Ria}^{-1}\)\(\sum\limits_{i=2}^3U_{Ria}^{-1*}U_{Ria}^{-1}+4U_{R1a}^{-1*}U_{R1a}\)\]\nn\\
&+&\f{1}{3}\sum\limits_{a=1}^5\(\sum\limits_{i=1}^3 Z_{Lia}^{-1*}Z_{Lia}^{-1}-\sum\limits_{i=2}^3 Z_{Ria}^{-1*}Z_{Ria}^{-1}\)\nn\\
&&\times \(\sum\limits_{i=1}^3 Z_{Lia}^{-1*}Z_{Lia}^{-1}-2\sum\limits_{i=4}^5 Z_{Lia}^{-1*}Z_{Lia}^{-1}-\sum\limits_{i=2}^3Z_{Ria}^{-1*}Z_{Ria}^{-1}+2\sum\limits_{i=1,4,5}Z_{Ria}^{-1*}Z^{-1}_{Ria}\) \nn\\
&+&\f{2}{9}\sum\limits_{a=1}^5\ln{\tl{\la}_a^2}\[ \(\sum\limits_{i=1}^3 Z_{Lia}^{-1*}Z_{Lia}^{-1}\)\(\sum\limits_{i=1}^3 Z_{Lia}^{-1*}Z_{Lia}^{-1}-2\sum\limits_{i=4}^5 Z_{Lia}^{-1*}Z_{Lia}^{-1}\)\.\nn\\&+&\left.\(\sum\limits_{i=2}^3 Z_{Ria}^{-1*}Z_{Ria}^{-1}\)\(\sum\limits_{i=1}^3Z_{Ria}^{-1*}Z_{Ria}^{-1}-2\sum\limits_{i=4}^5Z_{Ria}^{-1*}Z_{Ria}\)\]\nn\\
&+&\sum\limits_{a=1}^3\sum\limits_{b=1;a\neq b}^3\[\(\sum\limits_{i=1}^3 U_{Lia}^{-1*}U_{Lib}^{-1}\)\(\sum\limits_{i=1}^3 U_{Lib}^{-1*}U_{Lia}^{-1}\)\.\nn\\
&&+ \left.\(\sum\limits_{i=2}^3 U_{Ria}^{-1*}U_{Rib}^{-1}\)\(\sum\limits_{i=2}^3U_{Rib}^{-1*}U_{Ria}^{-1}+4U_{R1b}^{-1*}U_{R1a}\)\]\(\f{1}{3}-4 P(\la_a,\la_b)\) \nn\\
&+&\sum\limits_{a=1}^5\sum\limits_{b=1;a\neq b}^5\[\(\sum\limits_{i=1}^3 Z_{Lia}^{-1*}Z_{Lib}^{-1}\)\(\sum\limits_{i=1}^3 Z_{Lib}^{-1*}Z_{Lia}^{-1}
-2\sum\limits_{i=4}^5 Z_{Lib}^{-1*}Z_{Lia}^{-1}\)\.\nn\\
&&+\left. \(\sum\limits_{i=2}^3 Z_{Ria}^{-1*}Z_{Rib}^{-1}\)\(\sum\limits_{i=2}^3Z_{Rib}^{-1*}Z_{Ria}^{-1}-2\sum\limits_{i=1,3,4}U_{Rib}^{-1*}U_{Ria}\)\]\(\f{1}{3}-4 P(\tl{\la}_a,\tl{\la}_b)\) \nn\\
&+&\sum\limits_{a=1}^5\sum\limits_{b=1;a\neq b}^5\[\(\sum\limits_{i=1}^3 Z_{Lia}^{-1*}Z_{Lib}^{-1}\)\(\sum\limits_{i=2}^3Z_{Rib}^{-1*}Z_{Ria}^{-1}-2\sum\limits_{i=1,3,4}U_{Rib}^{-1*}U_{Ria}\)\.\nn\\
&&+\left. \(\sum\limits_{i=2}^3 Z_{Ria}^{-1*}Z_{Rib}^{-1}\)\(\sum\limits_{i=1}^3 Z_{Lib}^{-1*}Z_{Lia}^{-1}
-2\sum\limits_{i=4}^5 Z_{Lib}^{-1*}Z_{Lia}^{-1}\)\]2\tl{\la}_a\tl{\la}_b Q(\tl{\la}_a,\tl{\la}_b) \nn\\
&+&\sum\limits_{a=1}^3\sum\limits_{b=1;a\neq b}^3\[\(\sum\limits_{i=1}^3 U_{Lia}^{-1*}U_{Lib}^{-1}\)\(\sum\limits_{i=2}^3U_{Rib}^{-1*}U_{Ria}^{-1}+4U_{R1b}^{-1*}U_{R1a}\)\.\nn\\
&&+ \left.\(\sum\limits_{i=2}^3 U_{Ria}^{-1*}U_{Rib}^{-1}\)\(\sum\limits_{i=1}^3 U_{Lib}^{-1*}U_{Lia}^{-1}\)\]2\la_a\la_b Q(\la_a,\la_b)~.
\eeqa
\normalsize
with the definition
\beqa
P(a,b)&=&\int\limits_{0}^1 x(x-1)\ln[(a-b)x+b] dx~,\nn\\
Q(a,b)&=&\int\limits_{0}^1 \f{x(x-1)}{(a-b)x+b} dx~,
\eeqa
which we will not give their tedious analytic expressions.
\end{acknowledgments}

\end{document}